\documentclass[11pt]{article}

\usepackage[a4paper,margin=2.5cm]{geometry}
\usepackage[T1]{fontenc}
\usepackage[utf8]{inputenc}
\usepackage{lmodern}
\usepackage{microtype}
\usepackage{amsmath}
\usepackage{graphicx}
\usepackage[font=small,labelfont=bf]{caption}
\usepackage{booktabs}
\usepackage{multirow}
\usepackage{float}
\usepackage{needspace}
\usepackage[numbers,sort&compress]{natbib}
\usepackage{authblk}
\usepackage[hidelinks]{hyperref}
\title{Staged Hybridisation for Visual Quantum Reinforcement Learning via Knowledge Distillation}

\author[1,2]{Javier Lazaro}
\author[1]{Juan-Ignacio Vazquez}
\author[1]{Pablo Garcia-Bringas}

\affil[1]{University of Deusto, Bilbao, Spain}
\affil[2]{FSAS International Quantum Center, Fujitsu}

\affil[]{\texttt{javier.lazaro@opendeusto.es}, \texttt{ivazquez@deusto.es}, \texttt{pablo.garcia.bringas@deusto.es}}

\date{}

\begin{document}
\maketitle

\begin{abstract}
Visual environments are a demanding setting for quantum reinforcement learning (QRL): high-dimensional observations, unstable RL optimisation, and constrained variational quantum circuits (VQCs) are difficult to train jointly. This paper studies knowledge distillation (KD) as a staged hybridisation strategy for visual QRL. Instead of training a hybrid visual agent end-to-end from pixels, we first train a classical visual teacher, freeze its encoder as a feature interface, and distil the teacher's policy behaviour into compact downstream heads. These heads can be classical or VQC-based, enabling small quantum-compatible students to be evaluated under the same frozen representation as compact classical controls.

We evaluate the pipeline on \texttt{CartPole Pixels} and \texttt{Acrobot Pixels}. The results show that staged KD enables shallow VQC heads to acquire non-trivial visual-control behaviour in settings where direct pixel-based training would be substantially more difficult. Angle-encoded VQC heads retain near-teacher performance, while amplitude-encoded heads push compactness to an extreme regime, at the cost of greater fragility, stronger budget sensitivity, and higher simulation time. Overall, staged KD reframes visual QRL as a compact-head learning problem, opening a practical route for training small quantum-compatible policies outside the standard end-to-end RL loop.
\end{abstract}



\section{Introduction}
\label{sec:introduction}

Visual environments are a natural stress test for quantum reinforcement learning (QRL), and among its least forgiving settings. Variational quantum circuits (VQCs) have been explored as trainable components inside hybrid QRL agents, with promising results in low-dimensional control tasks and growing interest across the field~\citep{meyerSurveyQuantumReinforcement2024,alomariSurveyQuantumReinforcement2025a}. Yet pixel-based control changes the problem qualitatively: the agent must learn perception, control, and a quantum-compatible representation within the same unstable RL loop.

A direct end-to-end route is therefore a poor fit for small near-term VQC models. The issue is not only whether a particular circuit is expressive enough, but whether the training setup asks the quantum component to solve the right part of the problem. In visual domains, a shallow VQC is ill-suited to absorb the full perceptual burden from raw pixels. A more practical route is to separate perception from compact control.

This paper studies knowledge distillation (KD) as a staged hybridisation strategy for visual QRL. We first train a classical visual teacher, freeze its encoder, and then distil the teacher's policy behaviour into compact downstream heads. These heads can be classical or VQC-based, allowing small quantum-compatible students to be trained over the same frozen feature interface as compact classical controls. The goal is to test whether staged KD can turn a high-dimensional visual control problem into a simpler learning setting, where shallow VQC-based heads become viable.

We evaluate the pipeline on \texttt{CartPole Pixels} and \texttt{Acrobot Pixels}. The results show that compact classical heads and angle-based VQC heads retain most of the teacher's visual-control behaviour, while amplitude-based VQC heads provide a more extreme but fragile compactness condition. We further show that downstream KD budget is a key variable: some heads are stable with limited distilled experience, while others require substantially more data to become usable. Overall, the paper argues that staged training is a practical entry point for studying visual QRL with small and shallow VQC components.


\section{Related Work}
\label{sec:related_work}

QRL was first introduced as a theoretical framework for combining quantum information processing with reinforcement learning \citep{dongQuantumReinforcementLearning2008}. More recent work has moved toward practical hybrid agents based on VQCs, but this shift has also exposed persistent issues in trainability, architectural sensitivity, and fair comparison against classical baselines. Recent benchmarking work therefore calls for more controlled evaluation protocols in QRL \citep{meyerBenchmarkingQuantumReinforcement2025,kruseBenchmarkingQuantumReinforcement2025}. Component-wise studies reach a similar conclusion: the behaviour of hybrid QRL agents depends on the interaction between embedding, ansatz design, measurement, and post-processing, not on the quantum block in isolation \citep{lazaroDissectingQuantumReinforcement2025a}.

These issues become sharper in visual environments. Lockwood and Si explored hybrid quantum--classical agents on Atari, but reported that their models failed to learn meaningful policies on \texttt{Breakout} and \texttt{Pong} \citep{lockwoodPlayingAtariHybrid2021}. More recently, Freinberger et al. showed that Atari-style visual control can be addressed with a quantum--classical model, but through a pipeline that still relies on classical feature encoding and post-processing around the VQC \citep{freinbergerQuantumclassicalReinforcementLearning2024}. Nagy et al. take a related route by learning a classical latent observation space before training a quantum agent on the compressed representation \citep{nagyHybridQuantumclassicalReinforcement2025}. Together, these works suggest that visual QRL is less about replacing an entire visual policy with a VQC, and more about designing the interface between perception and quantum-compatible control.

Knowledge distillation provides one way to build such an interface. Classical KD transfers the behaviour of a larger model into a smaller student using softened targets \citep{hintonDistillingKnowledgeNeural2015b}, and policy distillation extends this idea to RL agents \citep{rusuPolicyDistillation2016}. In quantum machine learning, related transfer and teacher--student methods have been used to combine pre-trained classical representations with small quantum models \citep{mariTransferLearningHybrid2020a}, transfer knowledge from classical networks to QNNs \citep{hasanBridgingClassicalQuantum2025a}, and compress larger QNNs into smaller quantum architectures \citep{yanDistillingKnowledgeQuantum2026,alamKnowledgeDistillationQuantum2023}. These works motivate KD as a tool for quantum-compatible compression, but they do not address visual QRL.

Adjacent work in quantum imitation and offline QRL also departs from pure online trial-and-error training \citep{chengQuantumImitationLearning2023,eisenmannModelbasedOfflineQuantum2024}. However, these approaches do not address the visual setting considered here, where the main difficulty is the interface between high-dimensional perception and a constrained quantum-compatible control module. We therefore study a staged alternative: train a visual teacher first, freeze its encoder, and distil its policy behaviour into compact classical and VQC-based heads over the resulting feature-level representation.


\section{Staged Hybridisation via Knowledge Distillation}
\label{sec:staged_hybridisation}

The pipeline follows directly from this staged view of visual QRL. Instead of asking a hybrid agent to learn perception, policy optimisation, and VQC-based control in a single end-to-end loop, we separate these roles. A classical teacher first learns the pixel-based task; its visual encoder is then frozen as a stable feature interface; and KD is used to train compact downstream heads over that interface.

\begin{figure}[H]
    \centering
    \includegraphics[width=0.95\textwidth]{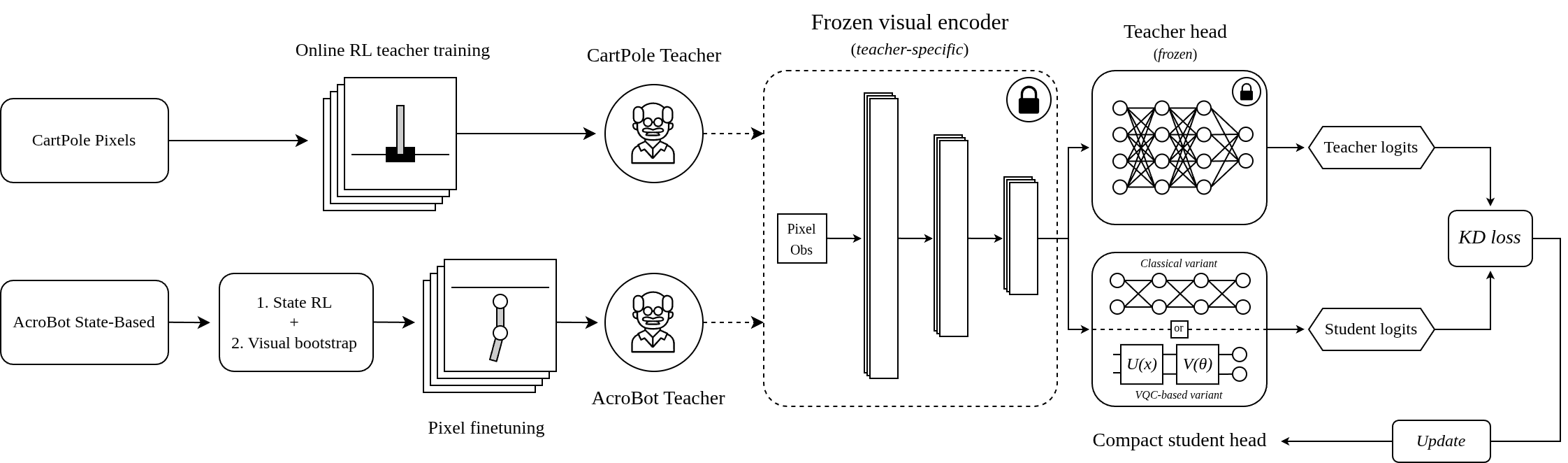}
    \caption{Overview of the staged hybridisation pipeline for visual QRL. A visual teacher is trained first, directly for \texttt{CartPole Pixels} and through state-based RL, visual bootstrap, and pixel-space finetuning for \texttt{Acrobot Pixels}. The teacher-specific encoder is then frozen as a feature interface, and offline KD trains compact classical or VQC-based student heads from teacher policy logits.}
    \label{fig:methodology_pipeline}
\end{figure}

Figure~\ref{fig:methodology_pipeline} summarises the proposed pipeline. The teacher provides both the visual representation and the policy targets, while the student learns only the downstream control behaviour from feature-level examples. The student head can be classical or VQC-based, which allows shallow quantum-compatible heads to be evaluated under the same frozen representation as compact classical controls. In this setting, KD is not simply a compression tool; it is the mechanism that turns visual control into a compact-head learning problem.

\paragraph{Stage 1: visual teacher construction.}
We begin by treating each environment as a classical online RL task and train a visual teacher capable of producing reliable behaviour from pixel observations. The difficulty of this teacher-construction stage depends on the environment. In dense-reward settings, such as \texttt{CartPole Pixels}, a direct visual RL teacher can be obtained from stacked image observations. In harder settings, such as \texttt{Acrobot Pixels}, the teacher may require an additional staged route before reaching a competent visual policy, combining state-based RL, visual bootstrap, and pixel-space finetuning. This upstream staging is part of the methodological point of the paper: visual control problems need not be approached as a single end-to-end optimisation problem, even before introducing a quantum student.

\paragraph{Stage 2: frozen visual interface and distillation data.}
Once a teacher has been obtained, we decompose it into a visual encoder and a downstream policy head. The visual encoder is retained as a frozen, teacher-specific feature extractor, while the teacher head provides the policy targets to be distilled. We then roll out the trained teacher in the environment and export an offline feature-level distillation dataset. For each visited pixel observation $o_i$, the frozen encoder produces a latent representation
\[
    \phi_i = f_T(o_i),
\]
and the teacher head provides the corresponding policy logits $z_i^T$ and selected action $a_i^T$. For the distillation experiments considered in this work, the relevant dataset can therefore be written as
\[
    \mathcal{D}_{KD}
    =
    \{(\phi_i, z_i^T, a_i^T)\}_{i=1}^{N}.
\]

This dataset defines the fixed interface on which all students are trained. The student no longer learns directly from raw pixels; instead, it receives the representation produced by the frozen teacher encoder and learns to reproduce the teacher head's behaviour on that representation. The amount and diversity of teacher-generated experience therefore become part of the downstream training regime. As shown in the experimental analysis, this is particularly relevant for highly compact students, where limited coverage of the teacher's state distribution can affect the final policy.

\paragraph{Stage 3: compact-head distillation.}
The student replaces the original teacher head with a compact trainable head $h_S$. Given a frozen teacher feature vector $\phi_i$, the student produces logits
\[
    z_i^S = h_S(\phi_i).
\]
The distillation target is the teacher policy distribution, following the general principle of policy distillation~\citep{rusuPolicyDistillation2016}. In our implementation, the student is trained with a temperature-scaled KL-divergence loss between teacher and student logits:
\[
    \mathcal{L}_{KD}
    =
    T^2
    \frac{1}{N}
    \sum_{i=1}^{N}
    D_{\mathrm{KL}}
    \left(
        \mathrm{softmax}\left(z_i^T / T\right)
        \,\middle\|\,
        \mathrm{softmax}\left(z_i^S / T\right)
    \right),
\]
where $z_i^T$ are the teacher logits and $T$ is the distillation temperature. This objective trains the student to imitate the teacher's softened policy distribution over actions, rather than only the hard action selected by the teacher. Although this work focuses on policy-logit distillation, the same staged formulation could naturally be extended to value distillation, joint policy-value distillation, or other teacher-student objectives.

\paragraph{Student head families.}
The staged formulation is architecture-agnostic with respect to the downstream head. This allows us to compare compact classical heads and VQC-based heads under the same frozen visual representation and the same distillation protocol. In this work, the student families include compact classical heads, angle-encoded VQC heads, and amplitude-encoded VQC heads. The VQC-based students are deliberately small and shallow, reflecting the constraints that motivate the staged regime in the first place. Angle-encoded heads represent the main practical hybrid variant, while amplitude-encoded heads provide a more aggressively compact but potentially fragile alternative.

The compactness concerns the trainable downstream head, not the full deployed pipeline, as the frozen visual encoder remains part of the final student agent. This is central to the proposed methodology: the goal is not to compress the entire visual policy into a quantum model, but to transfer the control behaviour of a visual teacher into a small trainable head that can be evaluated as part of a frozen-encoder student policy.

\paragraph{Final deployment and evaluation.}
After distillation, the original teacher head is discarded. The final student agent is formed by composing the frozen teacher encoder with the trained compact student head,
\[
    \pi_S(o)
    =
    \arg\max_a
    \left[
        h_S(f_T(o))
    \right]_a .
\]
This composed policy is then evaluated directly in the original pixel-based environment. The evaluation therefore tests whether the compact student head has acquired enough of the teacher's visual-control behaviour to act successfully from image observations.

The methodological goal of this pipeline is to make small-head visual control a viable training regime for VQC-based QRL. By moving the visual representation burden into a frozen teacher encoder, shallow VQC heads no longer need to learn directly from pixels or rely on large end-to-end CNN+VQC architectures. Instead, they can be trained as compact policy heads over a stable visual interface. The same interface also supports compact classical students, which serve both as strong controls and as evidence that the proposed staging is a general training strategy whose relevance is especially acute for quantum models constrained by circuit depth, trainability, and input dimensionality.


\Needspace{15\baselineskip}
\section{Experimental Design}
\label{sec:experimental_design}

This section describes how the staged KD pipeline is instantiated experimentally. The main goal is to evaluate whether compact classical and VQC-based student heads can acquire visual-control behaviour from frozen teacher representations. We therefore focus the experimental design on three elements: the construction of competent visual teachers, the definition of the frozen feature interface and student-head families, and the evaluation regimes used to compare downstream compact heads.

\subsection{Environments and teacher construction}
\label{subsec:envs_teachers}

We evaluate the proposed pipeline on two pixel-based control benchmarks: \texttt{CartPole Pixels} and \texttt{Acrobot Pixels}. Both environments use visual observations rendered at $84 \times 84$ resolution with four stacked frames. The two tasks play complementary roles. \texttt{CartPole Pixels} provides a relatively accessible visual-control setting with dense rewards, while \texttt{Acrobot Pixels} acts as a more informative stress test due to its harder dynamics and sparser reward structure.

Teacher construction follows the first stage described in Section~\ref{sec:staged_hybridisation}. For \texttt{CartPole Pixels}, competent teachers can be obtained through direct online visual RL. For \texttt{Acrobot Pixels}, obtaining a strong visual teacher from reward alone was itself non-trivial; the retained teachers were therefore produced through a staged route involving state-based RL, visual bootstrap, and pixel-space finetuning. This difference is part of the experimental design: it allows us to test the staged KD pipeline both in a setting where visual teacher construction is relatively direct and in a setting where staging is already required before student distillation.

Table~\ref{tab:selected_teachers} summarises the retained teacher checkpoints used to generate the downstream KD datasets. These teachers define the frozen visual interfaces and policy targets used in all student-head experiments.

\begin{table}[H]
    \centering
    \caption{Retained teacher checkpoints used for downstream KD. CartPole uses direct visual teachers, while Acrobot uses bootstrap-finetuned visual teachers. Mean return corresponds to the retained teacher checkpoint used for downstream distillation. The final row in each block reports the mean and standard deviation across the five retained teachers.}
    \label{tab:selected_teachers}
    \begin{minipage}[t]{0.48\textwidth}
        \centering
        \small
        \textbf{CartPole Pixels}\\[0.4em]
        \setlength{\tabcolsep}{4pt}
        \begin{tabular*}{\linewidth}{@{\extracolsep{\fill}}cc}
            \toprule
            Seed & Mean return \\
            \midrule
            0 & 497.1 \\
            1 & 479.0 \\
            2 & 482.8 \\
            3 & 500.0 \\
            4 & 493.8 \\
            \midrule
            Mean $\pm$ SD & $490.5 \pm 9.2$ \\
            \bottomrule
        \end{tabular*}
    \end{minipage}
    \hfill
    \begin{minipage}[t]{0.48\textwidth}
        \centering
        \small
        \textbf{Acrobot Pixels}\\[0.4em]
        \setlength{\tabcolsep}{4pt}
        \begin{tabular*}{\linewidth}{@{\extracolsep{\fill}}cc}
            \toprule
            Seed & Mean return \\
            \midrule
            0 & -77.5 \\
            1 & -74.2 \\
            2 & -78.8 \\
            3 & -66.4 \\
            4 & -73.3 \\
            \midrule
            Mean $\pm$ SD & $-74.0 \pm 4.8$ \\
            \bottomrule
        \end{tabular*}
    \end{minipage}
\end{table}

\subsection{Frozen interface and student head families}
\label{subsec:student_heads}

After teacher construction, each teacher is split into a frozen visual encoder and a downstream policy head. The frozen encoder is used to export a feature-level KD dataset, as described in Section~\ref{sec:staged_hybridisation}. All students trained from the same teacher replica therefore receive the same 512-dimensional teacher feature interface and are optimised against the same teacher policy targets. This design isolates the downstream control head as the object of comparison.

We evaluate five student-head variants, reported as \textbf{Classical (D512)}, \textbf{Classical (D16)}, \textbf{Classical (D8)}, \textbf{Angle VQC (8Q)}, and \textbf{Amplitude VQC (9Q)}. The three classical students share the same basic structure: an optional trainable linear bottleneck, followed by a one-hidden-layer MLP with 64 hidden units and a final linear policy layer. Classical (D512) uses the full frozen representation directly, without a bottleneck. Classical (D16) and Classical (D8) first project the 512-dimensional teacher representation to 16 or 8 dimensions, respectively, before applying the same MLP policy head.

The two VQC-based students replace the classical MLP head with a shallow quantum-compatible head followed by a linear readout. In both cases, the VQC has two variational layers. Each layer applies trainable $R_Y$ and $R_Z$ rotations on every qubit, followed by a ring of CNOT entangling gates. The circuit returns one Pauli-$Z$ expectation value per qubit, and a final linear layer maps these expectation values to the action logits required by the environment.

Angle VQC (8Q) uses a trainable linear bottleneck from the 512-dimensional frozen representation to an 8-dimensional pre-VQC interface. These eight values are scaled and embedded through $Y$-axis angle encoding into an 8-qubit circuit. This variant represents the main practical hybrid line: it keeps the VQC small and shallow while still allowing a learned classical interface between the frozen visual encoder and the quantum circuit. The 8-dimensional interface was selected as the smallest bottleneck that remained sufficiently stable in preliminary classical checks; more aggressive compression made the downstream policy noticeably less reliable.

Amplitude VQC (9Q) removes the explicit trainable bottleneck and instead uses amplitude encoding to load the full 512-dimensional frozen representation into the quantum state. In amplitude encoding, the input vector is mapped to the amplitudes of the computational basis states. Since an $n$-qubit state has $2^n$ amplitudes, 9 qubits are sufficient to represent 512 input features. This makes the amplitude-based student an aggressively compact contrast condition, where the dimensional reduction is handled by the quantum state representation rather than by a learned bottleneck. The input is normalised as part of the amplitude-embedding operation before the same shallow variational circuit and linear readout are applied.

\subsection{Evaluation regimes and metrics}
\label{subsec:evaluation_metrics}

The evaluation is organised into two complementary regimes. The main comparison evaluates all five student-head variants under the full downstream KD dataset, providing a controlled comparison between compact classical heads and VQC-based heads under a shared training regime. The budget study evaluates a smaller set of variants, Classical (D8), Angle VQC (8Q) and Amplitude VQC (9Q), under reduced downstream budgets. This second regime tests how each family behaves as the amount of teacher-generated experience available for distillation is reduced.

For each run, students are evaluated periodically during KD training and again at the end of training. We report both the final checkpoint and the best checkpoint retained during the run. Let $R_{\mathrm{final}}$ denote the 100-episode evaluation return of the final checkpoint, and let $R_{\mathrm{best}}$ denote the 100-episode evaluation return of the best retained checkpoint. The selected return is defined as
\[
    R_{\mathrm{selected}}
    =
    \max
    \left(
        R_{\mathrm{final}},
        R_{\mathrm{best}}
    \right).
\]
This selected score is used as the main performance metric to capture agents that peak before the final checkpoint. Reporting only the final checkpoint can underestimate the best deployable student obtained during distillation. To make this behaviour explicit, we also report the observed drift between the selected and final results:
\[
    \Delta_{\mathrm{drift}}
    =
    R_{\mathrm{selected}}
    -
    R_{\mathrm{final}}.
\]
A drift close to zero indicates that the final checkpoint retains the best observed performance, while larger values indicate degradation after the best retained checkpoint.

In addition to return-based metrics, we report trainable head parameters, parameter reduction relative to the uncompressed Classical (D512) head within each environment, total training time, and QNode execution time for quantum-enabled variants. All VQC heads are evaluated in noiseless statevector simulation using the \texttt{lightning.qubit} backend. Timing measurements were obtained on a workstation with an AMD Ryzen 9 9900X CPU, 16 GB of RAM, and Ubuntu 24.04. Exact hyperparameters and runner configurations for each run are provided in the accompanying repository.

Overall, this design tests three questions: whether staged KD can transfer visual-control behaviour into compact downstream heads, whether small and shallow VQC-based heads can operate under the same frozen visual interface used by compact classical controls, and how sensitive the different student families are to downstream KD budget.

\section{Results}
\label{sec:results}

\subsection{Main comparison under a unified downstream budget}
\label{subsec:main_comparison}

We first evaluate all student-head variants under the full downstream KD dataset. This comparison tests how much of the teacher's visual-control behaviour is retained when the original teacher head is replaced by compact classical or VQC-based alternatives.

Table~\ref{tab:main_results} reports the main student return-based results under the full-budget regime, while the retained teacher references are reported separately in Table~\ref{tab:selected_teachers}. In \texttt{CartPole Pixels}, all compact classical heads retain near-teacher performance despite large downstream compression. Classical (D8) reaches a selected return of $488.5 \pm 10.4$, essentially matching Classical (D512) at $488.7 \pm 10.3$. The Angle VQC (8Q) head follows the same pattern, reaching $487.2 \pm 5.1$ with a shallow VQC-based head.

\begin{table}[H]
    \centering
    \caption{Main performance results under the full-budget regime. Returns are reported as mean $\pm$ standard deviation across replicas. Drift is reported as selected minus final return.}
    \label{tab:main_results}
    \small
    \setlength{\tabcolsep}{6pt}
    \begin{tabular*}{\textwidth}{@{\extracolsep{\fill}}llccc}
        \toprule
        Environment & Variant & Selected return & Final return & Drift \\
        \midrule
        \multirow{5}{*}{CartPole Pixels}
            & Classical (D512)   & $488.7 \pm 10.3$ & $486.3 \pm 11.0$ & $2.4$ \\
            & Classical (D16)    & $487.0 \pm 11.8$ & $482.1 \pm 12.2$ & $4.8$ \\
            & Classical (D8)     & $488.5 \pm 10.4$ & $486.0 \pm 11.1$ & $2.4$ \\
            & Angle VQC (8Q)     & $487.2 \pm 5.1$  & $483.2 \pm 8.8$  & $4.0$ \\
            & Amplitude VQC (9Q) & $396.7 \pm 59.4$ & $389.1 \pm 68.5$ & $7.6$ \\
        \midrule
        \multirow{5}{*}{Acrobot Pixels}
            & Classical (D512)   & $-74.8 \pm 4.7$ & $-75.3 \pm 5.0$ & $0.6$ \\
            & Classical (D16)    & $-74.9 \pm 4.8$ & $-75.4 \pm 5.3$ & $0.5$ \\
            & Classical (D8)     & $-74.4 \pm 4.5$ & $-75.2 \pm 5.2$ & $0.8$ \\
            & Angle VQC (8Q)     & $-75.8 \pm 2.8$ & $-78.1 \pm 3.1$ & $2.3$ \\
            & Amplitude VQC (9Q) & $-81.5 \pm 3.9$ & $-81.9 \pm 4.0$ & $0.5$ \\
        \bottomrule
    \end{tabular*}
\end{table}

The same pattern appears in \texttt{Acrobot Pixels}. The compact classical heads remain close to the retained visual teachers, with Classical (D8) obtaining the strongest selected return among the students at $-74.4 \pm 4.5$. The Angle VQC (8Q) remains close at $-75.8 \pm 2.8$, showing that a shallow VQC-based head can retain useful control behaviour even in the environment where teacher construction itself required a staged route.

The amplitude-encoded student is consistently the hardest full-budget regime. Amplitude VQC (9Q) learns useful behaviour, which is non-trivial in pixel-based control tasks, but its returns are lower and more variable than those of the other heads. In \texttt{CartPole Pixels}, it drops to $396.7 \pm 59.4$, while in \texttt{Acrobot Pixels} the degradation is milder but still visible. This supports its role as an extreme compactness condition: viable, but substantially more fragile than the angle-based variant.

Overall, the full-budget comparison shows that staged KD can transfer visual-control behaviour into compact downstream heads. The Angle VQC (8Q) is the strongest quantum-compatible variant under this regime, while compact classical heads remain the strongest controls. The remaining question is whether the fragility of the amplitude-based variant is fixed by the architecture alone, or whether it can be mitigated by increasing the amount of distilled teacher experience.

\subsection{Budget sensitivity as a methodological variable}
\label{subsec:budget_sensitivity}

We next analyse how the compact-head families behave when the downstream KD budget is reduced. Here, budget refers to the number of teacher-generated feature-level examples available for student distillation, not to the online RL budget used to construct the teacher.

\begin{figure}[H]
    \centering
    \includegraphics[width=0.95\textwidth]{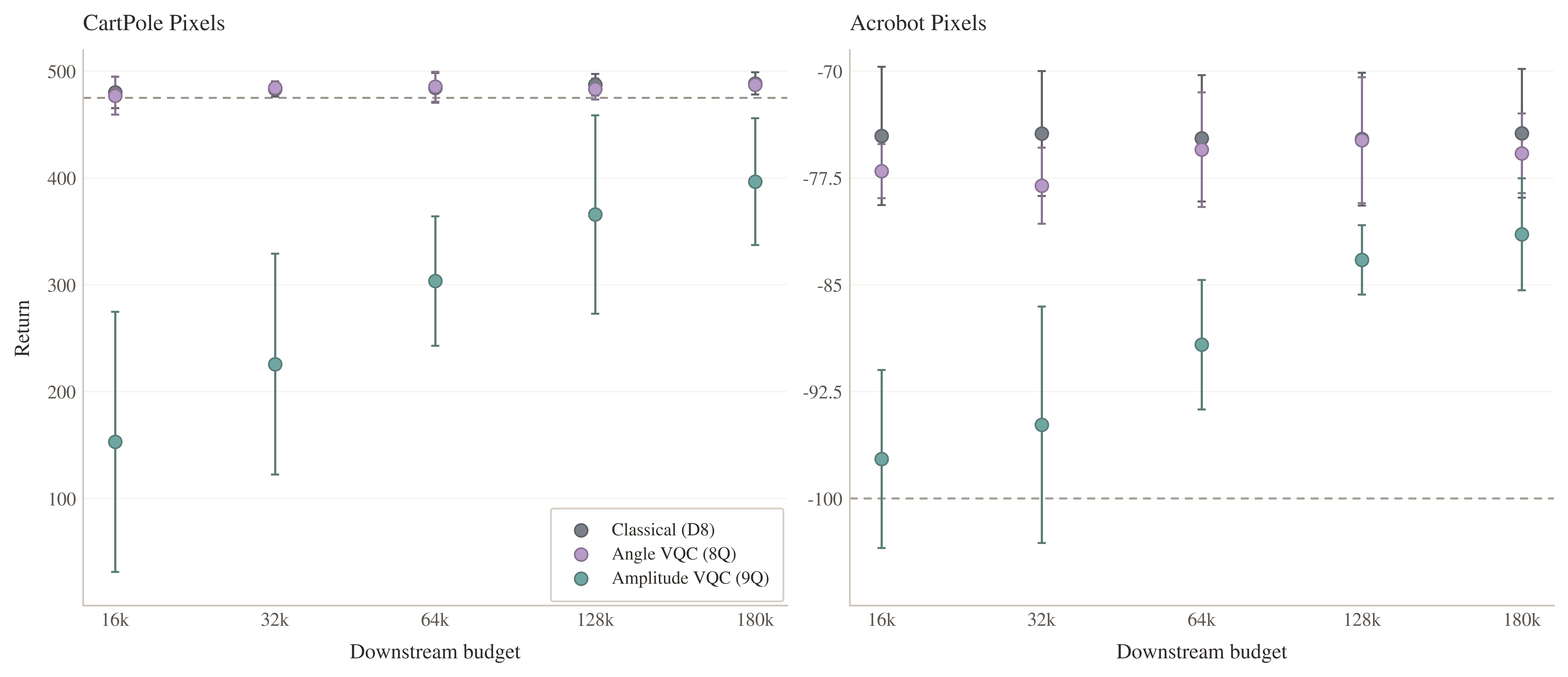}
    \caption{Budget sensitivity of the three compact downstream families. Each point reports selected return under a fixed downstream KD budget, using teacher-generated feature-level examples. Dashed lines indicate the solved threshold in each environment: 475 for \texttt{CartPole Pixels} and $-100$ for \texttt{Acrobot Pixels}.}
    \label{fig:budget_study}
\end{figure}

Figure~\ref{fig:budget_study} reports selected return for Classical (D8), Angle VQC (8Q), and Amplitude VQC (9Q) across downstream budgets. The compact classical head is largely insensitive to budget in both environments: it is already close to full-budget behaviour at 16k examples in \texttt{CartPole Pixels}, and remains stable across the sweep in \texttt{Acrobot Pixels}.

The Angle VQC (8Q) follows a similar but slightly more variable pattern. It reaches the solved regime in \texttt{CartPole Pixels} even at the smallest budget, and remains close to the compact classical control. In \texttt{Acrobot Pixels}, it is more variable at low and intermediate budgets, but approaches teacher-level behaviour as more downstream data are provided.

The strongest budget dependence appears in Amplitude VQC (9Q). In both environments, selected return improves consistently as the amount of distilled teacher experience increases. This trend is especially important because the amplitude-based student has no learned pre-VQC bottleneck: the shallow VQC operates directly on the amplitude-embedded frozen representation. The budget study therefore clarifies the full-budget result: amplitude encoding can acquire non-trivial visual-control behaviour, but it requires substantially more downstream data to do so reliably.

Taken together, the budget study shows that downstream sample budget changes the qualitative behaviour of the student families. Classical (D8) is stable across budgets, Angle VQC (8Q) remains the most robust quantum-compatible option, and Amplitude VQC (9Q) behaves as a viable but data-hungry extreme-compression regime. This makes downstream budget useful but not sufficient as an evaluation axis: it must be interpreted together with the parameter and runtime costs of each compact-head family.

\Needspace{12\baselineskip}
\subsection{Computational trade-offs}
\label{subsec:compute_tradeoffs}

We finally examine the computational trade-offs associated with each full-budget student family. Table~\ref{tab:compute_tradeoffs} reports trainable downstream-head parameters, parameter reduction relative to Classical (D512), total training time, and the fraction of time spent inside QNode execution for quantum-enabled variants.

\begin{table}[H]
    \centering
    \caption{Full-budget computational trade-offs. Training time is reported as the median across replicas. Parameter reductions are computed relative to the trainable downstream head of the Classical (D512) baseline within each environment.}
    \label{tab:compute_tradeoffs}
    \small
    \setlength{\tabcolsep}{4pt}
    \begin{tabular*}{\textwidth}{@{\extracolsep{\fill}}llcccc}
        \toprule
        Environment & Variant & Head params & Red. (\%) & Train (min) & QNode (\%) \\
        \midrule
        \multirow{5}{*}{CartPole Pixels}
            & Classical (D512)   & 32962 & --   & 5.9   & --   \\
            & Classical (D16)    & 9426  & 71.4 & 6.3   & --   \\
            & Classical (D8)     & 4810  & 85.4 & 6.3   & --   \\
            & Angle VQC (8Q)     & 4154  & 87.4 & 66.8  & 83.7 \\
            & Amplitude VQC (9Q) & 56    & 99.8 & 115.6 & 86.7 \\
        \midrule
        \multirow{5}{*}{Acrobot Pixels}
            & Classical (D512)   & 33027 & --   & 1.3   & --   \\
            & Classical (D16)    & 9491  & 71.3 & 1.3   & --   \\
            & Classical (D8)     & 4875  & 85.2 & 1.3   & --   \\
            & Angle VQC (8Q)     & 4163  & 87.4 & 53.7  & 88.0 \\
            & Amplitude VQC (9Q) & 66    & 99.8 & 103.7 & 88.4 \\
        \bottomrule
    \end{tabular*}
\end{table}

The trade-off is strongly architecture-dependent. Compact classical heads reduce trainable parameters by more than 85\% at D8 while preserving essentially the same training-time scale as the uncompressed Classical (D512) downstream head. This makes the compact classical variants strong controls: they show that the staged interface itself supports substantial local compression without requiring a VQC.

The VQC-based heads expose a different regime. Angle VQC (8Q) reaches a comparable trainable-parameter reduction to Classical (D8), and slightly higher reduction in both environments, while retaining near-teacher performance. This comes with a much larger simulation cost: training time rises from 6.3 to 66.8 minutes in \texttt{CartPole Pixels}, and from 1.3 to 53.7 minutes in \texttt{Acrobot Pixels}. In both cases, QNode execution accounts for most of the runtime, reaching 83.7\% and 88.0\%, respectively.

Amplitude VQC (9Q) pushes trainable-head compactness to the most extreme regime, with only 56--66 trainable parameters and 99.8\% reduction relative to Classical (D512). However, it is also the slowest variant and the least stable in performance, especially in \texttt{CartPole Pixels}. These results reinforce the interpretation used throughout the paper: staged KD makes compact quantum-compatible visual-control heads trainable and evaluable, but it does not make VQC simulation computationally inexpensive.

\begin{figure}[H]
    \centering
    \includegraphics[width=0.95\textwidth]{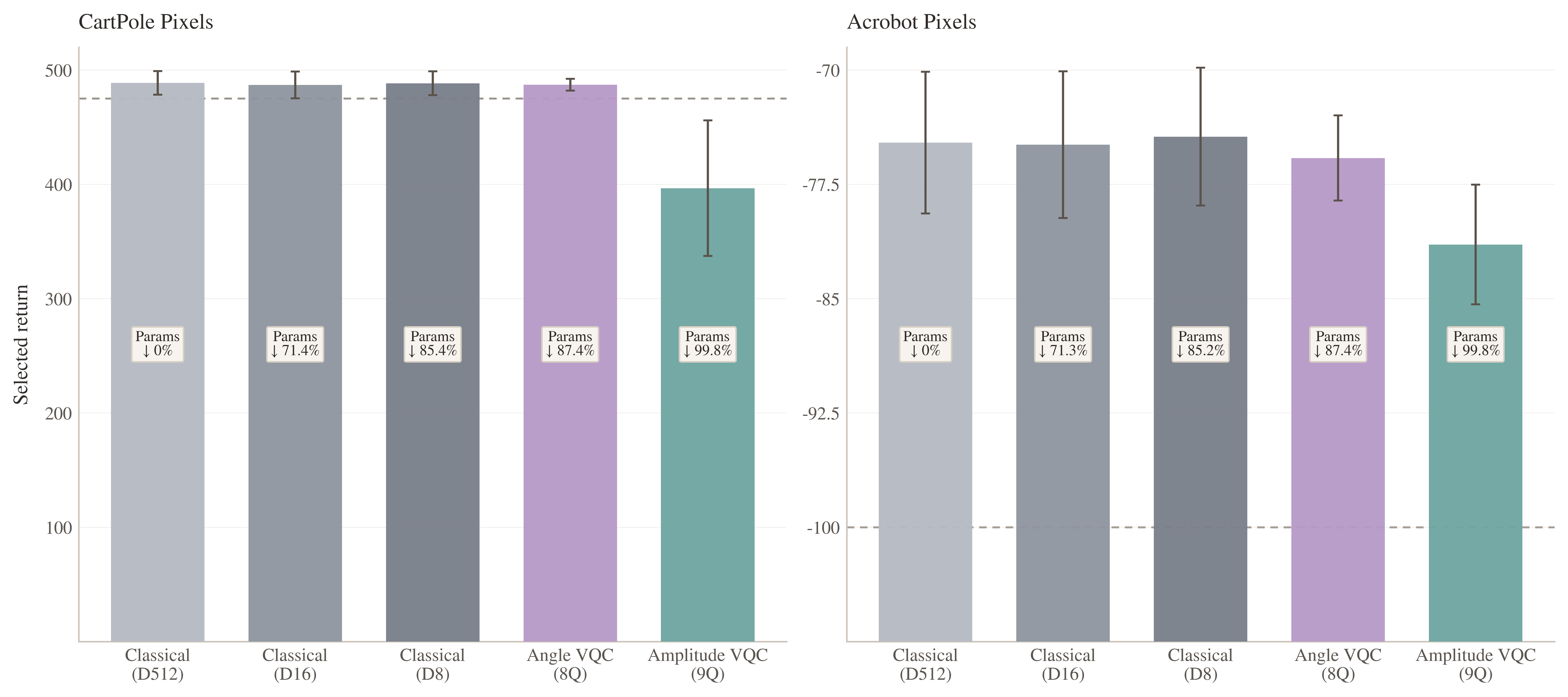}
    \caption{Full-budget trade-off between selected return and trainable-head parameter reduction relative to the Classical (D512) baseline.}
    \label{fig:parameter_performance_tradeoff}
\end{figure}

Figure~\ref{fig:parameter_performance_tradeoff} summarises the parameter--performance trade-off. Compact classical heads and Angle VQC (8Q) preserve most of the policy quality while strongly reducing the trainable downstream head. Amplitude VQC (9Q) achieves the largest parameter reduction, but at the cost of lower returns, stronger budget sensitivity, and higher simulation time.

The empirical picture is therefore consistent across the three analyses. Staged KD enables compact downstream heads, including shallow VQC-based heads, to retain non-trivial visual-control behaviour. However, the practical trade-off is architecture-dependent: angle encoding provides the most balanced hybrid regime, while amplitude encoding exposes the tension between extreme compactness, downstream data requirements, and quantum execution cost. These trade-offs are interpreted more broadly in the following discussion.


\section{Discussion}
\label{sec:discussion}

The results support the central methodological claim of this work: staged KD provides a practical route for training compact quantum-compatible control heads in visual RL tasks. The key finding is that shallow VQC-based students can acquire non-trivial pixel-based control behaviour once the perceptual burden has been absorbed by a frozen teacher encoder. This changes the experimental regime for VQC-based visual QRL: instead of asking a complex hybrid model to learn perception and control jointly from raw pixels, the quantum component is evaluated as a compact downstream policy head over a stable visual interface.

The strongest hybrid result is obtained by the angle-based VQC. Its performance remains close to the compact classical controls in both environments and across most downstream budgets. This suggests that the learned pre-VQC bottleneck acts as both a dimensionality-reduction device and as a useful classical-to-quantum interface. It allows the frozen teacher features to be reshaped before quantum processing, making the shallow VQC head a more practical component compared to the direct embedding route.

The amplitude-based VQC illustrates the opposite extreme. It achieves the largest reduction in trainable parameters by loading the 512-dimensional frozen representation into a 9-qubit amplitude state, but this compactness comes with lower returns, stronger budget sensitivity, and higher simulation cost. This does not make the amplitude variant uninformative. On the contrary, it acts as a stress test for extreme compactness: it shows that useful visual-control behaviour can still be distilled into a very small quantum-compatible head, but also that removing the learned pre-VQC interface makes the regime substantially more fragile.

A key limitation is that compactness is local rather than global. The frozen visual encoder remains part of the deployed student policy, so the method does not compress the entire visual agent into a quantum model. The claim is instead that the trainable downstream control head can be made compact and quantum-compatible while retaining useful behaviour. This distinction is essential for interpreting the results: the pipeline is a staged hybridisation strategy, not an end-to-end quantum replacement for visual RL.

Compact classical heads remain strong controls, matching or outperforming VQC-based heads in several instances. This is an important part of the empirical picture rather than a weakness of the setup: the staged pipeline makes it possible to compare compact classical and quantum-compatible heads under the same frozen visual interface. The value of the proposed methodology lies in providing a controlled way to study small VQC heads in visual-control settings where direct end-to-end QRL would be substantially harder to train and interpret. The results should therefore be read as evidence for a practical staged training regime for visual QRL, not as evidence that quantum heads are currently superior to classical counterparts.


\section{Conclusion}
\label{sec:conclusion}

This paper presented staged hybridisation via knowledge distillation as a practical route for training compact quantum-compatible control heads in visual reinforcement-learning tasks. Rather than training a visual QRL agent end-to-end from pixels, the proposed pipeline first trains a classical visual teacher, freezes its encoder, and then distils the teacher's policy behaviour into compact downstream heads. This separates the perceptual burden from the final control problem and creates a controlled regime in which small classical and VQC-based students can be evaluated under the same frozen visual interface.

Across \texttt{CartPole Pixels} and \texttt{Acrobot Pixels}, the results show that compact downstream heads can retain useful visual-control behaviour. The angle-based VQC head emerges as the most practical hybrid variant, remaining close to classical controls. The amplitude-based VQC provides an extreme compactness condition, demonstrating that useful behaviour can be distilled into a very small quantum-compatible model, but with greater fragility, stronger budget sensitivity, and higher simulation cost.

Overall, staged KD changes the experimental regime for visual QRL: it enables shallow VQC heads to be evaluated on visual-control tasks without training a full end-to-end pixel policy. Future work should use this regime as a warm start for partially end-to-end or on-policy hybrid training and extend it to richer visual environments and hardware-aware settings.

\section*{Code Availability}

Code, experiment configurations, hyperparameters, runner scripts, logged results, and plotting utilities required to reproduce the reported results are available in the accompanying repository: \href{https://github.com/javier-lazaro/visual-qrl-kd}{github.com/javier-lazaro/visual-qrl-kd}.

\section*{Acknowledgements}

The authors used OpenAI's ChatGPT for language editing and drafting assistance. All scientific content, experimental design, results, interpretations, and final manuscript decisions were reviewed and validated by the authors.

\newpage

\bibliographystyle{plainnat}
\bibliography{references}

\end{document}